\documentclass[12pt]{article}

\usepackage{mathrsfs}
\usepackage{amsmath}

\global\arraycolsep=1pt
\usepackage[colorlinks=true,backref=true,linkcolor=black,anchorcolor=black,citecolor=black,filecolor=black,menucolor=black,pagecolor=black,urlcolor=black]{hyperref}
\usepackage[Symbol]{upgreek}
\usepackage{bbm}
\usepackage{dsfont}
\usepackage{amssymb}
\usepackage{textcomp}

\newcommand{\scal}[1]{\bigl ({#1} \bigr )}

\newcommand{\CR}{\nonumber \\*}
\newcommand{\del}{\partial}

\newcommand{\trace}{\hbox {Tr}~}

\DeclareMathAlphabet{\mathpzc}{OT1}{pzc}{m}{it}

\def\Lc{\mathscr{L}}

\def\d{\mathfrak{d}}

\def\A{a}

\def\C{c}

\def\F{\mathscr{F}}

\usepackage[colorlinks=true,backref=true,linkcolor=black,anchorcolor=black,citecolor=black,filecolor=black,menucolor=black,pagecolor=black,urlcolor=black]{hyperref}

\def\d{\delta}

\def\gk{g(\kappa)}
\def\ik{i_\kappa}

\def\Idvt{\int {\rm d}\vte^m {\rm d}\vte^n\,}

\def\A{\mathbb{A}}
\def\F{\mathbb{F}}
\def\C{\mathbb{C}}
\def\Gag{\mathbf{\Upgamma}}
\def\PPsi{\mathbf{\Uppsi}}
\def\CChi{\boldsymbol{\upchi}}
\def\bCChi{\bar{\boldsymbol{\upchi}}}

\def\PPhi{\mathbf{\Upphi}}
\def\bPPhi{\bar{\mathbf{\Upphi}}}

\usepackage[Symbol]{upgreek}
\usepackage{caption}
\usepackage{bbm}
\usepackage{dsfont}
\usepackage{amssymb}
\usepackage{textcomp}

\def\Lc{\mathscr{L}}

\def\N{\mathcal{N}}

\def\d{\delta}

\def\ga{\gamma}
\def\te{\theta}
\def\vte{\vartheta}

\def\gauge{\delta^{\scriptscriptstyle \,gauge}}

\def\be{\begin{equation}}
\def\ee{\end{equation}}
\def\bea{\begin{eqnarray}}
\def\eea{\end{eqnarray}}
\def\bdis{\begin{displaymath}}
\def\edis{\end{displaymath}}
\def\nn{\nonumber}

\def\d{\delta}

\def\gk{g(\kappa)}
\def\ik{i_\kappa}

\def\nn{\nonu}
\def\bea{\begin{eqnarray}}
\def\eea{\end{eqnarray}}
\def\be{\begin{equation}}
\def\ee{\end{equation}}

\def\m{{\bar m}}
\def\n{{\bar n}}

\let\nonu=\nonumber

\begin{document}
\allowdisplaybreaks[1]
\renewcommand{\thefootnote}{\fnsymbol{footnote}}
\def\corr{$\spadesuit $}
\def\trefle{$\clubsuit$}

\renewcommand{\thefootnote}{\arabic{footnote}}
\setcounter{footnote}{0}


 \def\stop{$\blacksquare$}
\begin{titlepage}
\renewcommand{\thefootnote}{\fnsymbol{footnote}}
\begin{flushright}
\
\vskip -3cm
{\small CERN-PH-TH/2008-156}\\
\vskip 3cm
\end{flushright}
\begin{center}
{{\Large \bf
Holomorphic  Superspace
 }}
\lineskip .75em
\vskip 3em
\normalsize
{\large Laurent Baulieu\footnote{email address: baulieu@lpthe.jussieu.fr},
Alexis Martin\footnote{email address: alexis.martin@lpthe.jussieu.fr}\\
\vskip 1em
$^{* }${\it Theoretical Division CERN} \footnote{ CH-1211 Gen\`eve, 23, Switzerland}
\\
$^{* \dagger}${\it LPTHE, CNRS and Universit\'e Pierre et Marie
Curie } \footnote{ 4 place Jussieu, F-75252 Paris Cedex 05,
France}
\\
}

\vskip 1 em
\end{center}
\vskip 1 em
\begin{abstract}

We   give a   twisted  holomorphic superspace  description  for the   super-Yang--Mills theory,    using   holomorphic and antiholomorphic  decompositions   of  twisted spinors.  We consider the case  of the $\N=1$ super-Yang--Mills theory in four dimensions.
 We solve the constraints in two different manners, without and with a prepotential. This might have further application for an holomorphic  superspace description  of $\N=1,d=10$ theory. We also explain how   the $\N=1$ and  $\N=2$   holomorphic superspaces are related.

\end{abstract}

\end{titlepage}
\renewcommand{\thefootnote}{\arabic{footnote}}
\setcounter{footnote}{0}

\section*{Introduction}

  The construction of a superspace path integral formulation for maximal supersymmetry is still an open question. To get a supersymmetry algebra that admits a functional representation on the fields is at the heart of the problem and it seems inevitable in dimensions $d\geqslant 7$ that this implies a breaking of the manifest Lorentz invariance.\footnote{In lower dimensions, there is still the possibility to formulate the maximally supersymmetric YM theory in terms of a subalgebra of the whole super-Poincar\'e algebra, while maintaining manifest Lorentz invariance. In $d=4$ for instance, a harmonic superspace formulation was given preserving $3/4$ of the supersymmetries \cite{N=3}, which then received a full quantum description \cite{Delduc}.} 

 Such a functional representation was determined   in \cite{10DSYM} for the $\N=1,d=10$ theory,    by  a supersymmetry algebra made of $9$ generators and  a restriction of the ten-dimensional Lorentz group to $SO(1,1)\times Spin(7)\subset SO(1,9)$. This   led   us    to a   reduced superspace with  $9$ fermionic coordinates. Covariant constraints were found, which do not imply   equations of motion. They  were solved in function  of the fields of the component formalism and
 analogous   results have been obtained for  the  $\N=2,d=4,8$ cases \cite{twistsp}.
 
 A path integral formulation was given for $\N=2,d=4$ in terms of the connection superfields themselves, which required an implementation of the constraints directly in the path integral. On the other hand, dimensional arguments show that the introduction of a prepotential is needed  in the higher dimensional cases. Moreover, we expect such higher dimensional cases to be formulated in terms of complex representations of $SU(4)\subset Spin(7)$. 
Using an  $SU(4)$ holomorphic formulation in 8 or 10 dimensions implies  a framework  that is formally ``similar'' to the holomorphic formulation in four dimensions that we study in this paper.  

We thus   display a holomorphic superspace formulation of the simple $\N=1,d=4$ super-Yang--Mills theory in its twisted form, by applying  the  general procedure of \cite{twistsp}. This superspace formulation involves $3$ supercharges, a scalar and a $(1,0)$-vector. It completes the previous works for the $\N=2,\ d=4$ and $\N=1,\ d=10$ twisted superspace with $5$ and $9$ supercharges, respectively. We also  provide a short discussion of the resolution of the constraints in terms of a prepotential.  It must be noted that reality conditions are a delicate issue for  the $\N=1\  d=4 $ superspace in holomorphic coordinates. However, this question does not arise in the 10-dimensional formulation, so we will not discuss it here.

The first section defines  the notations  of the holomorphic $\N=1,d=4$ super-Yang--Mills theory  and its formulation in components. The second  gives its superspace formulation together with the coupling to    matter. The third  provides     a discussion on the alternative formulation in superspace involving a prepotential. The fourth section   is devoted to the $\N=2$ case, both in components and superspace formulations.

\section { Holomorphic $\N=1,d=4$  Yang--Mills  supersymmetry }

\def\m{ {\bar m}}
\def\n{ {\bar n}}

The twist procedure for the $\N=1,d=4$ super--Yang--Mills theory has been described in \cite{Johansen,Witten1,recons} in the context of topological field theory. For a hyperK\"ahler manifold, one can use a pair of covariantly constant spinors $\zeta_{\pm}$,    normalized by $\zeta_{-\dot{\alpha}}\zeta^{\dot{\alpha}}_+=1$. They can be defined by $iJ^{m \bar n}\sigma_{m\bar n}\zeta_{\pm}=\pm\zeta_{\pm}$, where $J^{m\bar n}$ is the complex structure
\be
\label{J}
J^{mn} = 0 \ , \quad\quad J^{\m\n}=0 \ , \quad\quad J^{m\bar n} = i g^{m\bar n}
\ee
It permits one to decompose forms into holomorphic and antiholomorphic components. For the gauge connection $1$-form $A$, one has
\be
A = A_{(1,0)} + A_{(0,1)}  \quad \textrm{with} \quad JA_{(1,0)} = i A_{(1,0)} \ ,  JA_{(0,1)} = - i A_{(0,1)}
\ee
and the decomposition of its curvature is $F=dA+AA=F_{(2,0)}+F_{(1,1)}+F_{(0,2)}$.
A Dirac spinor  decomposes as
\be
\lambda_\alpha=\Psi_m\,\sigma^m_{\alpha\dot{\alpha}}\,\zeta_-^{\dot{\alpha}}\qquad \lambda^{\dot{\alpha}}=\eta\,\zeta_+^{\dot{\alpha}}+\chi_{\bar m\bar n}\,\sigma^{\bar m\bar n\dot{\alpha}}_{\phantom{\bar m\bar n\dot{\alpha}}\dot{\beta}}\,\zeta^{\dot{\beta}}_+
\ee
In the case of a flat manifold, the twist  is a mere   rewritting of the   Euclidean supersymmetric theory, obtained by mapping all spinors onto ``holomorphic'' and ``antiholomorphic'' forms after reduction of the $Spin(4)$ covariance to $SU(2)$. Notice that the Euclidean formulation of the   $\N=1$ theory     is defined as the analytical continuation of the Minkowski theory. The Euclideanization procedure  produces   a doubling of the fermions \cite{Nicolai.1978}, so that the complex fields      $\eta, \chi_{\bar m\bar n},\Psi_m$      are  truly mapped onto a Dirac spinor $\lambda$.  However, the  twisted and untwisted actions do not depend on the complex conjugate fields and the path integral can be   defined  as counting   only   four real degrees of freedom\footnote{In the Euclideanization procedure, one also gives up hermiticity of the action, but a ``formal complex conjugation'' can be defined and extended in the twisted component formalism that restores hermiticity \cite{recons}}. The twist also maps the four  $\N=1$  supersymmetry generators onto a (0,0)-scalar $\delta$, a (0,1)-vector $\delta_{\bar m}$  and a (2,0)-tensor  $\delta_{mn}$ generators. For    formulating the ``holomorphic superspace", we will only retain 3 of the    four   generators,   the scalar one  $\delta$ and the vector one  $\delta_{\bar m}$.  The invariance under $\delta$ and   $\delta_{\bar m}$  has been shown to completely determine the supersymmetric action \cite{recons}. Moreover, the absence of anomaly for the tensor symmetry implies that this property can be conserved at the quantum level (at least at any given finite order in perturbation theory) \cite{TSSlong}.

\subsection{Pure $\N=1$ super--Yang--Mills theory} \label{pure N1 components}
The bosonic fields content of the $\N=1$ pure super--Yang--Mills theory is made of the Yang--Mills field $A= A_m dz^m +A_\m  dz^\m $, and an auxiliary scalar field $T$, while the fermionic fields are one scalar $\eta$, one $(1,0)$-form $\Psi_m$ and one $(0,2)$-form $\chi_{\bar m\bar n}$.
The   transformation laws of the various fields in twisted representations are
\be \label{lois N1 d4}\begin{split}
\d\,A_m &= \Psi_m\\
\d\,A_{\bar m} &= 0\\
\d\,\Psi_m &= 0\\
\d\,\eta &= T\\
\d\,T &= 0\\
\d\,\chi_{\bar m\bar n} &= F_{\bar m\bar n}
\end{split} \hspace{10mm}
\begin{split}
\d_{\bar m}\,A_n &= g_{\bar m n}\eta\\
\d_{\bar m}\,A_{\bar n} &= \chi_{\bar m\bar n}\\
\d_{\bar m}\,\Psi_n &= F_{\bar m n}-g_{\bar m n} T\\
\d_{\bar m}\,\eta &= 0\\
\d_{\bar m}\,T &= D_{\bar m} \eta\\
\d_{\bar m}\,\chi_{\bar p\bar q} &= 0
\end{split}
\ee
The three equivariant generators  $\delta$ and  $\delta_{\bar m}$ verify the following off-shell supersymmetry algebra
\be\label{com}
\delta^2=0\, ,\quad \{\delta,\d_{\bar m}\}=\partial_{\bar m}+\gauge(A_{\bar m})\, , \quad \{\d_{\bar m},\d_{\bar n}\}=0
\ee
The  action for the  pure $\N=1,d=4$ super--Yang--Mills is completely determined by the $\delta,\delta_{\bar m}$ invariance. It is given by \cite{Johansen,recons}
\be \label{action N1 d4}
\mathcal{\,S}_{\,YM}^{\N=1} = \int\! \mathrm{d}^4x\sqrt{g}\,\trace\Bigl(\frac{1}{2}F^{mn}F_{mn}+T(T+iJ^{m\bar n}F_{m\bar n})-\chi^{mn}D_m\Psi_n+\eta D^m\Psi_m\Bigr)
\ee

The Wess and Zumino matter multiplet and its coupling to pure $\N=1$ super--Yang-Mills will     only  be discussed in the framework of superspace.

\subsection{Elimination of gauge transformations in the closure relations} \label{Monemvasia}
The algebra (\ref{com}) closes on gauge transformations, due to the fact that in superspace, where supersymmetry is linearly realized, one breaks the super-gauge invariance to get the transformation laws of the component fields (\ref{lois N1 d4}). To be consistent with supersymmetry, this in turn implies to modify the supersymmetry transformations by adding field dependent gauge transformations, resulting in non linear transformation laws. This super-gauge is analogous to the Wess and Zumino gauge in ordinary superspace and such an algebra is usually referred as an algebra of the Wess and Zumino type. In this section, we show  how the use  of shadow fields makes it possible to remove these gauge transformations,  by applying the general formalism of~\cite{ TSSlong} to the $\N=1,d=4$ case. This in turn permits one to make contact with the general solution to the superspace constraints given in the next section.

To introduce the shadows, one  replaces  the knowledge of the $\delta, \delta_{\bar m}$ generators by that of  graded differential operators $Q$ and $Q_\kappa$,   which  represent supersymmetry in a nilpotent way. Let $\omega$ and $\kappa^{\bar m}$ be the commuting  scalar and   $(0,1)$-vector supersymmetry parameters, respectively.  The actions of $Q$ and $Q_\kappa$ on the  (classical)  fields are   basically   supersymmetry transformations as in      (\ref{lois N1 d4}) minus a field dependent gauge transformation, that is
\be \label{def Q}
Q\equiv \omega\delta -\gauge(\omega c)\, ,\quad Q_\kappa\equiv\delta_\kappa -\gauge(\ik\ga_1)
\ee
with $\delta_\kappa\equiv \kappa^{\bar m}\delta_{\bar m}$ and $\ik$ is the contraction operator along $\kappa^{\bar m}$. These operators obey $Q^2=0,Q_\kappa^2=0,\{Q,Q_\kappa\}=\omega\mathcal{L}_\kappa$. The scalar shadow field $c$ and   the    $(0,1)$-form    shadow field    $\ga_1$    are a generalization of the fields introduced in \cite{shadow}. They carry a $U(1)_{\scriptstyle{R}}$ charge $+1$ and  $-1$, respectively. The action of $Q$ and  $Q_\kappa$ increases it by  $1$ and $-1$, respectively. Let moreover $\mathcal{Q}\equiv Q+Q_\kappa$. The property $\mathcal{Q}^2=\omega\mathcal{L}_\kappa$ fixes the transformation laws of $c$ and $\ga_1$. 
In fact, the action of  $\mathcal{Q}$         on all fields, classical and shadow ones,
is given by the following horizontality equation
\be \label{horizN1}
(d+\mathcal{Q}-\omega\ik)(A+\omega c+\ik\ga_1) +(A+\omega c+\ik\ga_1)^2=F+\omega\Psi_{(1,0)}+\gk\eta +\ik\chi
\ee
together with its Bianchi identity
\be
(d+\mathcal{Q}-\omega\ik)(F+\omega\Psi_{(1,0)}+\gk\eta +\ik\chi)+[A+\omega c+\ik\ga_1,F+\omega\Psi_{(1,0)}+\gk\eta +\ik\chi]=0
\ee
implied by $(d+\mathcal{Q}-\omega\ik)^2=0$. Here and elsewhere $g(\kappa)\equiv g_{m \bar m}\kappa^{\bar m} dz^m$. The transformation laws (\ref{lois N1 d4})  can indeed be recovered from  these horizontality equations by expansion over form degree and $U(1)_{\scriptstyle{R}}$ number, modulo      gauge transformations  with parameters
             $\omega c$ or  $   \ik\ga_1$. The auxiliary $T$ scalar field is introduced  in order to solve the degenerate equation involving      $Q   \,  g(\kappa) \eta  +Q_\kappa \,\omega  \Psi$, with $Q \eta = \omega T -[\omega c,\eta]$. Moreover, the fields in the r.h.s of (\ref{horizN1}) can be interpreted as curvature components.

             Let us turn to the action of $\mathcal{Q}$ on the shadow fields.
              For the sake of
             notational  simplicity, we will omit from now on the dependence on  the scalar parameter  $\omega$. To recover its dependence, it is sufficient to remember that  $Q$ increases the $U(1)_{\scriptstyle{R}}$ number by one unit. The horizontality conditions imply three equations for the shadow fields
\be
Qc=-c^2\, ,\quad Q(\ik\ga_1) +Q_\kappa c +[c,\ik\ga_1]= \ik A\, ,\quad Q_\kappa (\ik\ga_1)=-(\ik\ga_1)^2
\ee
Due to the nilpotency of $\ik$, the third equation is defined modulo a contracted $(0,2)$ even form $\ga_2$ of $U(1)_{\scriptstyle{R}}$ number $-2$, that is $Q_\kappa\ga_1 =i_\kappa\ga_2+\frac{1}{2}[\ga_1,\ik\ga_1]$. To  solve the second equation, we introduce an odd $(0,1)$-form $c_1$ of $U(1)_{\scriptstyle{R}}$ number zero. This gives  $Q\ga_1 = c_1-[c,\ga_1]$ and $Q_\kappa c =\ik c_1 +\ik A$. Since we must have   $\mathcal{Q}^2=\mathcal{L}_\kappa$   on all fields,  we  find
\be \begin{split}
     Q\ga_1 &= c_1-[c,\ga_1]\\
     Q\ga_2 &= c_2-[c,\ga_2]-\frac{1}{2}[c_1,\ga_1]
    \end{split}\hspace{10mm}\begin{split}
Q c &= -c^2\\
Q c_1 &= -[c,c_1]\\
Q c_2 &= -[c,c_2]-c_1^2
\end{split}\ee
and
\be \begin{split}
     Q_\kappa\ga_1 &=i_\kappa\ga_2+\frac{1}{2}[\ga_1,\ik\ga_1]\\
     Q_\kappa \ga_2 &= \frac{1}{2}[\ga_1,\ik\ga_2]-\frac{1}{12}[\ga_1,[\ga_1,\ik\ga_1]]
    \end{split}\hspace{10mm}\begin{split}
Q_\kappa c &=\ik c_1 +\ik A\\
Q_\kappa c_1 &=\ik c_2+\Lc_\kappa\ga_1\\
Q_\kappa c_2 &= \Lc_\kappa\ga_2 -\frac{1}{2}[\ga_1, \Lc_\kappa\ga_1]
\end{split}\ee
with $\Lc_\kappa\equiv[\ik,d_A]$. 

\section{$\N=1,\ d=4$ holomorphic superspace}
\subsection{Definition of holomorphic superspace}

We now  define       a   `` twisted  holomorphic "  superspace  for $\N=1$ theories   by  extending
 the $z_m$, $z_{\bar{m}}$ bosonic space with three Grassmann coordinates, one scalar $\te$ and two $(0,1)$ $\vte^{\bar{p}}$ ($m,\bar{p}=1,2$).
The supercharges are given by
\bea \label{susy gen} \mathbb{Q}\,\,
\equiv
\frac{\partial}{\partial\te}+\vte^{\bar{m}}\partial_{\bar{m}},
\hspace{20mm} \mathbb{Q}_{\bar{m}}
\equiv\frac{\partial}{\partial\vte^{\bar m}} \CR
\mathbb{Q}^2=0,\hspace{10mm} \{\mathbb{Q},\mathbb{Q}_{\bar m}\}=
\partial_{\bar m}, \hspace{10mm} \{\mathbb{Q}_{\bar m}, \mathbb{Q}_{\bar n}\}=0
\eea
The covariant superspace derivatives and their anticommuting
relations are \bea \nabla\,\, \equiv \frac{\partial}{\partial\te}
\hspace{20mm} \nabla_{\bar m}
\equiv\frac{\partial}{\partial\vte^{\bar m}}-\te\partial_{\bar m}
\CR \nabla^2=0 \hspace{10mm} \{\nabla,\nabla_{\bar m}\}=
-\partial_{\bar m} \hspace{10mm} \{\nabla_{\bar m}, \nabla_{\bar
n}\}=0 \eea They anticommute with the supersymmetry generators.
They can be gauge-covariantized by the introduction of connection superfields $\mathcal{A}  \equiv  (\mathbb{C},\mathbf{\Upgamma}_{\bar m},\A_m,\mathbb{A}_{\bar m})$ 
valued in the adjoint of the gauge group of the theory   
\be \hat{\nabla} \equiv \nabla +
\mathbb{C},\quad \hat{\nabla}_{\bar m} \equiv \nabla_{\bar
m}+\mathbf{\Upgamma}_{\bar m},\quad \hat{\partial}_m\equiv
\partial_m+\mathbb{A}_m,\quad \hat{\partial}_{\bar m}\equiv
\partial_{\bar m}+\mathbb{A}_{\bar m} \ee 
The associated   covariant
superspace  curvatures are defined as ($M=m,\bar{m}$)
  \be \label{ricci}\begin{split}
&\mathbb{F}_{MN} \equiv [\hat{\partial}_M,\hat{\partial}_N] \\
&\mathbf{\Uppsi}_M \equiv [\hat{\nabla},\hat{\partial}_M] \\
&\boldsymbol{\upchi}_{\bar{m}N} \equiv
[\hat{\nabla}_{\bar m},\hat{\partial}_N]
\end{split}
\hspace{10mm}\begin{split}
&\mathbf{\Upsigma} \equiv \hat{\nabla}^2 \\
&\mathbb{L}_{\bar m} \equiv \{\hat{\nabla}, \hat{\nabla}_{\bar m}\}+\hat{\partial}_{\bar m} \\
&\mathbf{\bar{\Upsigma}}_{\bar m \bar n} \equiv
{\ \scriptstyle \frac{1}{2} } \{\hat{\nabla}_{\bar m},\hat{\nabla}_{\bar n} \}
\end{split}\ee
so that \be \label{curv}\begin{split}
\mathbb{F}_{M N} &= \partial_M\mathbb{A}_N -\partial_N\mathbb{A}_M +[\mathbb{A}_M,\mathbb{A}_N] \\
\mathbf{\Uppsi}_M &= \nabla \mathbb{A}_M - \partial_M\mathbb{C} - [\mathbb{A}_M,\mathbb{C}] \\
\boldsymbol{\upchi}_{\bar{m} N} &= \nabla_{\bar m}\mathbb{A}_N - \partial_N \mathbf{\Upgamma}_{\bar m}
- [\mathbb{A}_N,\mathbf{\Upgamma}_{\bar m}]
\end{split}
\hspace{10mm}\begin{split}
\mathbf{\Upsigma} &=\nabla\mathbb{C}+\mathbb{C}^2 \\
\mathbb{L}_{\bar m} &= \nabla\mathbf{\Upgamma}_{\bar m} +\nabla_{\bar m}\mathbb{C}+\{\mathbf{\Upgamma}_{\bar m}, \mathbb{C}\}+\mathbb{A}_{\bar m} \\
\mathbf{\bar{\Upsigma}}_{\bar m \bar n} &= \nabla_{\{ \bar m} \mathbf{\Upgamma}_{\bar n\}}
+ \mathbf{\Upgamma}_{\{\bar m} \mathbf{\Upgamma}_{\bar n\}}
\end{split}\ee
Bianchi identities are given by  $\Delta  \mathcal{F} =-[
\mathcal{A} , \mathcal{F} ]
  $, where $\Delta$ and $\mathcal{F}$ denote collectively $(\nabla,\nabla_{\bar m},\partial_m,\partial_{\bar m})$ and the superspace curvatures. The  super-gauge transformations of the super-connection $\mathcal{A}$ and super-curvature $\mathcal{F}$  are  \be
\label{gauge_transf N1} \mathcal{A} \rightarrow
e^{-\boldsymbol{\upalpha}}(\mathbf{\Updelta}+\mathcal{A})e^{\boldsymbol{\upalpha}},
\quad \mathcal{F} \rightarrow
e^{-\boldsymbol{\upalpha}}\mathcal{F}e^{\boldsymbol{\upalpha}} \ee
where the gauge superparameter $\boldsymbol{\upalpha}$ can be any
given general superfield valued in the Lie algebra of the gauge
group.  The  ``infinitesimal" gauge transformation is
$\delta \mathcal{A} =\mathbf{\Updelta} \boldsymbol{\upalpha} +[\mathcal{A} ,\boldsymbol{\upalpha} ]$.

\subsection{Constraints and their resolution}
The superfield interpretation of shadow fields is that they
parametrize the general $\boldsymbol{\upalpha}$-dependance of the
solution of   the superspace    constraints, while in components
they provide differential operators with no gauge transformations
in their anticommutation relations. To eliminate superfluous
degrees of freedom and   make contact with the component
formulation, we must impose the following gauge covariant
superspace constraints \be \label{constraints N1}
\mathbf{\Upsigma}= \mathbf{\bar{\Upsigma}}_{\bar m \bar
n}=\mathbb{L}_{\bar m} = 0,\quad \boldsymbol{\upchi}_{\bar m n} =
\frac{1}{2}g_{\bar m n}\boldsymbol{\upchi}^p_{\phantom{p}p}\equiv
g_{\bar m n}\boldsymbol{\eta}
 \ee
They can be solved in terms of component fields as follows.
The super-gauge symmetry (\ref{gauge_transf N1})
allows one to choose a super-gauge so that every antisymmetric as
well as the first component of $\Gag_{\bar m}$ is set to zero. We also fix the first
component $\C|_0=0$, so that we are left with the ordinary gauge
degree of freedom corresponding to $\boldsymbol{\upalpha}|_0$. The
constraint $\mathbf{\bar{\Upsigma}}_{\bar m \bar n}=0$ then implies
that the whole $\Gag_{\bar m}$ super-connection is zero.  The
constraint $\mathbf{\Upsigma}=0 $ implies that one must have \be
\C=\tilde{A}-\te\tilde{A}^2, \quad \tilde{A}|_0=0 \ee where $
\tilde{A} $ is a function of the $\vte_\m$. One defines
$(\frac{\del}{\del\vte^{\bar m}}\tilde{A})|_0\equiv - A_{\bar m}$.
The  constraint $\mathbb{L}_{\bar m} = 0$ implies \be \A_{\bar m}
= - \nabla_{\bar m} \C=-\frac{\del}{\del\vte^{\bar
m}}\tilde{A}+\te\Bigl(\del_{\bar
m}\tilde{A}-\frac{\del}{\del\vte^{\bar m}}\tilde{A}^2\Bigr) \ee
Then, with $(\frac{\del}{\del\vte^{\bar
m}}\frac{\del}{\del\vte^{\bar n}}\tilde{A})|_0\equiv \chi_{\bar
m\bar n}$, we have \be \CChi_{\bar m\bar n}=\nabla_{\bar
m}\A_{\bar n}=-\chi_{\bar m\bar n}-\te F_{\bar m\bar n} \ee It
follows that \be \C=-\vte^{\bar m}A_{\bar m}-\frac{1}{2}\vte^{\bar
m}\vte^{\bar n}\chi_{\bar m\bar n}-\te\Bigl(\frac{1}{2}\vte^{\bar
m}\vte^{\bar n}[A_{\bar m},A_{\bar n}]\Bigr) \ee It    only
remains to determine the field component content of $\A_m$. We
define $\A_m|_0\equiv A_m$, $(\frac{\del}{\del\te}\A_m)|_0\equiv
\Psi_m$ and $\boldsymbol{\eta}|_0\equiv \eta$. The   trace
constraint on $\CChi_{\bar m n}= \nabla_{\bar m}\A_n$ implies \be
\A_m = A_m +\vte^{\bar p}g_{\bar p m}\eta
+\te\Bigl(\Psi_m-\vte^{\bar p}(\del_{\bar p}A_m+g_{\bar p
m}T)+\vte^{\bar p}\vte^{\bar
q}g_{m[\bar{p}}\del_{\bar{q}]}\eta\Bigr)
\ee We  see that the whole physical content in the component
fields stand in the $\te$ independant part of the 
curvature superfield $\PPsi_m$, \be \PPsi_m|_{\te=0} = \Psi_m +\vte^{\bar
p}(F_{\bar p m}-g_{\bar p m}T)+\frac{1}{2}\vte^{\bar p}\vte^{\bar
q}(2 g_{m[\bar p}D_{\bar q]}\eta-D_m\chi_{\bar p\bar q}) \ee 
The general solution to the constraints can be obtained by a
super-gauge transformation, whose superfield parameter has
vanishing first component. It can be parametrized in various
manners. The following one allows one to recover the
transformation laws that we computed in components in the section (\ref{Monemvasia}) for the full set of
fields, including the scalar  and vectorial shadows \be
e^{\boldsymbol{\upalpha}}=e^{\te\vte^{\bar m}\partial_{\bar
m}}e^{\tilde{\gamma}}e^{\te\tilde{c}}= e^{\tilde{\gamma}}
\scal{1+\te( \tilde{c}+e^{-\tilde{\gamma}}\vte^{\bar
m}\partial_{\bar m} e^{\tilde{\gamma}})} \ee where $\tilde{\ga}$
and $\tilde{c}$ are respectively commuting and anticommuting
functions of $\vte^{\bar m}$ and the coordinates $z^m,z^{\bar m}$,
with the condition $\tilde{\ga}\vert_0=0$. These fields
appear here as the longitudinal degrees of freedom in superspace.
The transformation laws given in Eqs.~(\ref{lois N1 d4}) are
recovered for $\tilde{\gamma}=\tilde{c}=0$, modulo field-dependent
gauge-restoring  transformations.

\subsection{Pure $\N=1,d=4$ super-Yang--Mills action}

To express
the  pure super--Yang--Mills action in the twisted superspace,  we   observe that the Bianchi
 identity $\nabla\Psi_m +[\C,\Psi_m]$ implies that the gauge invariant
 function $\trace\Psi_m\Psi_n$ is $\te$ independent. Its component in $\vte^{\bar m}\vte^{\bar n}$ can
 thus be used to write an equivariant action as an integral over the full superspace
\begin{multline} \label{actionEQ} \mathcal{S}_{EQ} = \Idvt\, \trace
\Bigl(\PPsi_m\PPsi_n\Bigr)= \Idvt {\rm d}\te\,\trace\Bigl(
\A_m \, \PPsi_n - \C\del_m\A_n \Bigr)\\= \Idvt\,\mathrm{d}\te\,\trace\Bigl(
\A_m \,\nabla \A_n  -\C\F_{m n}\Bigr) \end{multline}
Berezin integration is defined as $\Idvt \mathbb{X}_{m n} \equiv -
\frac{1}{2}\frac{\partial}{\del\vte_{ m}}\frac{\partial}{\del\vte_{n}}\mathbb{X}_{m n}$,
where  $\mathbb{X}_{m n}$ is a     $(2,0)$-form superfield.
By use of the identity
$\trace(-\frac{1}{2}F^m_{\phantom{m}n}
F^n_{\phantom{m}m}+\frac{1}{2}F^m_{\phantom{m}m}F^n_{\phantom{m}n})=
\trace(\frac{1}{2}F_{m n}F^{m n}) + \textrm{''surface term``}$,
one recovers after implementation of the constraints the twisted
form of the $\N=1$ supersymmetric Yang--Mills action (\ref{action
N1 d4}), up to a total derivative \cite{Galperin.1991}.

Here, the constraints (\ref{constraints N1}) have been  solved in terms of component fields without using a prepotential.  They must be implemented directly in the path integral when one quantizes the theory, which run over the unconstrained potentials. This is performed by the following superspace integral depending
on   Lagrange multipliers superfields
\be \label{actionC N1}
\mathcal{S}_C = \Idvt {\rm d}\te\,\Omega_{mn}\trace\Bigl(\bar{\mathbb{B}}\,\Upsigma +\bar{\mathbb{B}}^{\bar m\bar n}\,\Upsigma_{\bar m\bar n} +\bar{\mathbb{K}}^{\bar m}\,\mathbb{L}_{\bar m}+\bar{\mathbf{\Uppsi}}^{m\bar n}\,\upchi_{m\bar n}\Bigr)
\ee
where $\bar{\mathbb{B}}^{\bar m\bar n}$ is symmetric and
$\bar{\mathbf{\Uppsi}}^{m\bar n}$ is traceless.
 The resolution of the constraints is such that the formal integration over the  above auxiliary superfields gives rise to the non-manifestly supersymmetric formulation of the theory in components, without introducing any determinant contribution in the path-integral. However, due to the Bianchi identities,   $\bar{\mathbb{B}}$, $\bar{\mathbb{B}}^{\bar m\bar n}$ and $\bar{\mathbf{\Uppsi}}^{m\bar n}$ admit a large class of zero
modes that must be considered in the manifestly supersymmetric superspace Feynman rules. They can be summarized by the following invariance of the action
\be \label{zero N1}
\delta^{\rm \scriptscriptstyle zero}\,\bar{\mathbb{B}}= \hat{\nabla}\,\mathbf{\uplambda}\, , \quad 
\delta^{\rm \scriptscriptstyle zero}\,
\bar{\mathbb{B}}^{\bar m\bar n} = \hat{\nabla}_{\bar p}\,
\mathbf{\uplambda}^{(\bar m\bar n\bar p)} - \partial_p\,
\mathbf{\uplambda}^{p\bar m\bar n}\, , \quad
\delta^{\rm \scriptscriptstyle zero}\, \bar{\mathbf{\Uppsi}}^{m\bar n} =
\hat{\nabla}_{\bar p}\, \mathbf{\uplambda}^{m \bar n\bar p} \ee
where $\mathbf{\uplambda}^{(\bar m\bar n\bar p)}$ is completely symmetric and $\mathbf{\uplambda}^{m \bar n\bar p}$ is traceless in its $m \bar n$ indices and symmetric in $\bar n\bar p$.
This feature is peculiar to twisted superspace and the appearance of this infinitely degenerated gauge symmetry was already underlined in \cite{twistsp} and is detailed in \cite{TSSlong}. We will not go in further details in this paper, and let the reader see in \cite{TSSlong} how it may be possible to deal with this technical subtelty by use of suitable projectors in superspace.

One    needs  a gauge fixing-action $\mathcal{S}_{GF}$. It is  detailed   for the analogous $\N=2$ twisted superspace in \cite{twistsp,TSSlong}   as    a    superspace generalization   of the   Landau gauge fixing action in components. One   also  needs a  gauge-fixing part $\mathcal{S}_{CGF}$ for the action of constraints (\ref{actionC N1}), and  the total action for $\N=1,d=4$ super-Yang--Mills in holomorphic superspace reads
\be
\mathcal{S}_{\tiny SYM}^{\tiny \N=1} = \mathcal{S}_{EQ}+\mathcal{S}_C +\mathcal{S}_{GF} + \mathcal{S}_{CGF}
\ee

\subsection{Wess and Zumino model}
We then turn to the matter content of the theory and consider as a first step the Wess and Zumino superfield formulation. We introduce two scalar superfields $\PPhi$ and $\bPPhi$, and one $(2,0)$-superfield $\bCChi_{m n}$. These superfields  correspond    to  the scalar chiral and anti-chiral superfields of ordinary  superspace. They take their values     in arbitrary representations of the gauge  group.   The   chirality constraints of the super-Poincar\'e  superspace are replaced  by  the following constraints
\be
\nabla\PPhi = 0, \quad \nabla_{\bar m}\bPPhi=0, \quad \nabla_{\bar p}\bCChi_{m n}=2\,g_{\bar p[m}\del_{n]}\bPPhi
\ee
We define the following component fields corresponding to the unconstrained components of the superfields as
$\bCChi_{m n}|_0\equiv \bar{\chi}_{m n}, (\frac{\del}{\del\te}\bCChi_{m n})|_0\equiv T_{m n},
\bPPhi|_0 \equiv\bar{\Phi}, (\frac{\del}{\del\te}\bPPhi)|_0\equiv\bar{\eta},
    \PPhi|_0 \equiv\Phi, (\frac{\del}{\del\vte^{\bar m}}\PPhi)|_0\equiv-\bar{\Psi}_{\bar m}, (\frac{\del}{\del\vte^{\bar m}}\frac{\del}{\del\vte^{\bar n}}\PPhi)|_0\equiv \bar{T}_{\bar m\bar n}$. We then deduce
\bea
\PPhi&=&\Phi-\vte^{\bar m}\bar{\Psi}_{\bar m}-\frac{1}{2}\vte^{\bar m}\vte^{\bar n}\bar{T}_{\bar m\bar n} \CR
\bPPhi&=&\bar{\Phi}+\te\Bigl(\bar{\eta}-\vte^{\bar m}\del_{\bar m}\bar{\Phi}\Bigr)\CR
\bCChi_{m n}&=&\bar{\chi}_{m n}+2\vte_{[m} \del_{n]}\bar{\Phi}+\te\Bigl(T_{m n}+\vte^{\bar m}(-\del_{\bar m}\bar{\chi}_{m n}+2g_{\bar m [m}\del_{n]}\bar{\eta})
+\vte_m\vte_n\del_{\bar p}\del^{\bar p}\bar{\Phi}\Bigr)
\eea
The free Wess and Zumino action can be written as
\bea
\mathcal{S}_{WZ} &=& \Idvt d\te\, \Bigl(-\PPhi\,\bCChi_{m n}\Bigr)\CR
 &=& \int\! \mathrm{d}^4x\sqrt{g}\,\trace\Bigl(\frac{1}{2}T^{\bar m\bar n}\bar{T}_{\bar m \bar n}-\chi^{\bar m\bar n}\partial_{\bar m}\bar{\Psi}_{\bar n}+\bar{\eta}\partial_m\bar{\Psi}^m-\bar{\Phi}\partial_m \partial^m\Phi\Bigr)
\eea

\subsection{Gauge coupling to matter}
In order to get the matter coupling to the pure super--Yang--Mills action, we covariantize the constraints. This can be shown to be consistent with those of (\ref{constraints N1}). We thus have
\be
\hat{\nabla}\PPhi = 0, \quad \hat{\nabla}_{\bar m}\bPPhi=0, \quad \hat{\nabla}_{\bar p}\bCChi_{m n}=2\,g_{\bar p[m}\hat{\del}_{n]}\bPPhi
\ee
In order to fulfil these new constraints, we modify the matter superfields as follows
\bea
\PPhi&=&\Phi-\vte^{\bar m}\bar{\Psi}_{\bar m}-\frac{1}{2}\vte^{\bar m}\vte^{\bar n}T_{\bar m\bar n}+\te\Bigl(\vte^{\bar m}A_{\bar m}\Phi+\vte^{\bar m}\vte^{\bar n}(\frac{1}{2}\chi_{\bar m\bar n}\Phi-A_{\bar m}\Psi_{\bar n})\Bigr)\CR
\bCChi_{m n}&=&\bar{\chi}_{m
n}+2\vte_{[m}D_{n]}\bar{\Phi}+\vte_m\vte_n\eta\bar{\Phi}
+\te\Bigl(T_{m n}+\vte^{\bar m}(-\del_{\bar m}\bar{\chi}_{m
n}+2g_{\bar m [m}D_{n]}\bar{\eta}\CR &&\hspace{50mm}-2g_{\bar m
[m}\Psi_{n]}\bar{\Phi})  +\vte_m\vte_n(\del_{\bar p}D^{\bar
p}\bar{\Phi}+\eta\bar{\eta}-h\bar{\phi})\Bigr) \eea The total
action of super--Yang--Mills coupled to matter then reads \be \label{Kithira}
\mathcal{S}_{SYM+Matter} = \Idvt d\te\, \Bigl(\trace(\A_m \,\nabla
\A_n  -\C\F_{m n})-\PPhi\,\bCChi_{m n}\Bigr) \ee 
which matches that of \cite{recons}. A WZ superpotential can be added in the twisted superspace formalism as the sum of two terms, one which is written as an integral over d$\te$ and the other as an integral over d$\vte^m$d$\vte^n$.

\section{Prepotential}
We now turn to the study of a twisted superspace formulation for the pure $\N=1$ super-Yang--Mills theory that involves a prepotential. It is sufficient to consider here the abelian case. The super-connections ($\C,\Gag_{\bar m},\mathbb{A}_{\bar m},\mathbb{A}_{m}$) count altogether for $(1+2+2+2)\cdot 2^3=56$ degrees of freedom, $8$ of which are longitudinal degrees of freedom associated to the gauge invariance in superspace (\ref{gauge_transf N1}). The constraints (\ref{constraints N1}) for $\Upsigma$ and $\Upsigma_{\bar m\bar n}$ can be solved by the introduction of unconstrained prepotentials as
\be
\mathbb{C} = \nabla \mathbb{D}\, , \quad  \Gag_{\bar m} = \nabla_{\bar m} \mathbf{\Updelta}
\ee
which reduces the $16$ degrees of freedom in $\Gag_{\bar m}$ to the $8$ degrees of freedom in $\mathbf{\Updelta}$. Gauge invariance for the prepotentials now reads
\be
\mathbb{D} \rightarrow \mathbb{D}+\upalpha\, ,\quad \mathbf{\Updelta} \rightarrow \mathbf{\Updelta}+\upalpha
\ee
Owning to their definition, the prepotentials are not uniquely defined. Indeed, they can be shifted by the additional transformations $\mathbb{D}\rightarrow \mathbb{D}+\mathbf{S}$ and $\mathbf{\Updelta} \rightarrow \mathbf{\Updelta}-\mathbf{T}$, where $\mathbf{S}$, $\mathbf{T}$ obey $\nabla \mathbf{S}=0$ respectively $\nabla_{\bar m}\mathbf{T}=0$. The constraint $\mathbb{L}_{\bar m}=0$ implies that $\mathbb{A}_{\bar m}$ can be expressed in terms of the prepotentials, $\mathbb{A}_{\bar m}=-\nabla\nabla_{\bar m}\mathbf{\Updelta} -\nabla_{\bar m}\nabla\mathbb{D}$. Its gauge invariance is given by
\be
 \mathbb{A}_{\bar m} \rightarrow -\nabla\nabla_{\bar m}(\mathbf{\Updelta} +\upalpha -\mathbf{T}) -\nabla_{\bar m}\nabla(\mathbb{D} +\upalpha +\mathbf{S}) = \mathbb{A}_{\bar m} +\partial_{\bar m}\upalpha
\ee We now perform a gauge choice and choose $\upalpha
=-\mathbf{\Updelta}$ so that we fix $\Gag_{\bar m}=0$. The
remaining gauge invariance is then given by $\mathbb{D}
\rightarrow \mathbb{D}+\mathbf{S}+\mathbf{T}$. The last constraint
$\boldsymbol{\upchi}_{\bar m n} =  g_{\bar m n}\boldsymbol{\eta}$
is then solved by introducing \be
\mathbb{A}_{m}=-\nabla^n\mathbb{P}_{nm} \ee
so that $\boldsymbol{\upchi}_{\bar m n}=\nabla_{\bar m}\mathbb{A}_{n}=\frac{1}{2}g_{\bar m n}\nabla_{\bar p}\nabla_{\bar q}\mathbb{P}^{\bar p\bar q}$. Although the residual gauge invariance is well established in the case when we consider the theory with the full set of generators, it is still unclear how exactly it happens in reduced superspace. But as a matter of fact, we are left with the two unconstrained prepotentials $\mathbb{D}$ and $\mathbb{P}_{mn}$, counting for $16$ degrees of freedom, which permits one to write the classical action. 
We consider the curvature $\PPsi_m$, which reads in terms of the prepotentials as
\be \label{Psim sf}
\PPsi_m = \nabla\nabla^n\mathbb{P}_{mn}-\partial_m\nabla\mathbb{D}
\ee
It obeys trivialy the Bianchi identity $\nabla\PPsi_m=0$, and its first component is a $(1,0)$-vector of canonical dimension $3/2$, that we identify with $\Psi_m-\partial_mc$ of the previous section, when the super-gauge invariance is restored.
   Since the first  component of the superfield (\ref{Psim sf}) is the same as that of $\PPsi_m$ in the previous section,   both superfields  are equal.
It follows that
the classical     $\N=1,d=4$ super-Yang--Mills action can also  be     written   as an integral over the full superspace,    in function  of an   unconstrained prepotential
\begin{multline} \label{actionEQ prep} \mathcal{S}_{EQ} = \Idvt\, \trace
\Bigl(\PPsi_m\PPsi_n\Bigr)= \Idvt {\rm d}\te\,\trace\Bigl(
\mathbb{P}_{mp}\nabla^p\nabla\nabla^q\mathbb{P}_{nq} + \mathbb{P}_{mp}\partial_n\nabla^p\nabla\mathbb{D}\Bigr)
\end{multline}
In fact, it       corresponds to the twisted version of the super-Poincar\'e  superspace action, once the tensorial coordinate  has been  eliminated.
We are currently studying how this formulation could be extended to a $\N=2,d=8$ superspace in a $SU(4)$ formulation \cite{TSSlong}.

\section{$\N=2 ,\, d=4$ holomorphic  Yang--Mills  supersymmetry}

%

We now define   the   holomorphic formulation of   the  $\N=2,\,d=4$   Yang--Mills  supersymmetry and see how it can be decomposed   into  that of     the  $\N=1$ supersymmetry.   We first focus on the component formulation and afterwards  we  give its superspace version. The latter will involve $5$ fermionic coordinates, as compared to the $3$ fermionic coordinates of the $\N=1$ twisted superspace.

\subsection{Component formulation}
The component formulation of $\N=2,d=4$ super-Yang--Mills in terms of complex representations has been discussed in \cite{Witten1, Marino.thesis, Park}. We consider a reduction of the euclidean rotation group $SU(2)_L\times SU(2)_R$ to $SU(2)_L\times U(1)_R$, with $U(1)_R\subset SU(2)_R$. The two dimensional representation of $SU(2)_R$ decomposes under $U(1)_R$ as a sum of one dimensional representations with opposite charges. In particular, the scalar and vector supersymmetry generators decompose as $\updelta=\delta +\bar\delta$ and $\updelta_{K}=\delta_\kappa +\bar\delta_{\bar \kappa}$, where $\bar \kappa$ is the complex conjugate of $\kappa$, so that $\vert\kappa\vert^2=i_{\bar \kappa}g(\kappa)$. The subsets $(\delta,\delta_\kappa)$ and  $(\bar\delta,\bar\delta_\kappa)$ form two $\N=1$ subalgebras of the  $\N=2$ supersymmetry, $(\delta,\delta_\kappa)$ being related to those of the previous sections
\bea
\delta^2=0\, ,\qquad \,\{\delta,\bar\delta\}&=&\gauge(\Phi)\, , \qquad \bar\delta^2=0\CR
\{\delta_m,\d_n\}=0\, ,\qquad \{\delta_m,\d_{\bar m}\}&=&g_{m\bar m}\gauge(\bar\Phi)\, ,\qquad \{\d_{\bar m},\d_{\bar n}\}=0
\eea
Quite concretely,   the  transformation laws for pure $\N=2$   super-Yang--Mills can be obtained from the  holomorphic and antiholomorphic decomposition of the  horizontality equation in $SU(2)\times SU(2)^{'}$ twisted formulation \cite{recons}. This equation involves the graded differential operator
\be
 {\cal Q} \equiv
 Q + Q_{K}
 \ee
which verify $(d +    Q + Q_K
  -  \varpi  i_{K})^2=0$. One defines 
 \be
 {\cal A}=  A_{(1,0)}
  +A_{(0,1)}
+ \varpi c
  +i_K\ga_{1}
  \ee
  where $c$ is a scalar shadow field and
  $\ga_1=\ga_{1\,(1,0)}+\ga_{1\,(0,1)}$  involves  ``holomorphic'' and ``antiholomorphic'' vector shadow fields. $Q$ and $Q_K$ are constructed out of the five $\updelta$, $\delta_\kappa$ and $\bar\delta_{\bar\kappa}$ supersymmetry generators with shadow dependent gauge transformations
\be \label{def Q N2}
Q\equiv \varpi\updelta -\gauge(\varpi c)\, ,\quad Q_K\equiv\delta_K -\gauge(i_K\ga_1)
\ee

An antiselfdual 2-form splits in holomorphic coordinates as $\chi \rightarrow (\chi_{(2,0)}, \chi_{(0,2)}, \chi_{(1,1)})$ where $\chi_{(1,1)}$ is subjected to the condition $\chi_{m\bar n}=\frac{1}{2}J_{m\bar n}J^{p\bar q}\chi_{p\bar q}$. We thus define a scalar $\chi$ as $\chi_{m\bar n}=g_{m\bar n}\chi$ and the holomorphic horizontality equation can be written as
 \bea\label{lois N2 d4}
 {\cal F }&\equiv&{\cal Q}  {\cal A}  +{\cal A} {\cal A} \nn\\
 &=&
 F_{(2,0)}
 + F_{(1,1)}
 + F_{(0,2)}
 +\varpi\Psi
  +g(\kappa) (\eta +\chi)   +g(\bar \kappa) (\eta -\chi) \CR
  &&\hspace{70mm}+i_{\bar \kappa}\chi_{(2,0)}
  +i_\kappa \chi_{(0,2)}
  +\varpi^2 \Phi
   +
    \vert\kappa\vert^2  \bar \Phi
  \eea
 with Bianchi identity
 \bea\label{Bianchi N2 d4}{
 \cal Q}       {\cal F }   =    -[{\cal A } ,
{\cal F }]
\eea
By expansion over form degrees and $U(1)_{\scriptstyle{R}}$ number, one gets transformation laws for $\N=2$ super--Yang--Mills in holomorphic and antiholomorphic components. In order to recover the transformation laws for $\N=1$ supersymmetry (\ref{lois N1 d4}) together with those of the matter multiplet in the adjoint representation $  \varphi = (T  _{\m\n}^{},   T_{mn } ^{},   \Psi_\m^{},    \chi_{mn}^{}    , \bar\eta,
\Phi^{},  \bar\Phi^ {}     )$, one can proceed as follows. One first derive from (\ref{def Q N2},\ref{lois N2 d4},\ref{Bianchi N2 d4}) the transformation laws for the physical fields under the equivariant operator $\delta_K$. One then obtains the action of the $\N=1$ vector generator by restricting the constant vector $K$ to its antiholomorphic component $\kappa^{\bar m}$, so that $\delta_K\rightarrow \kappa^{\bar m}\delta_{\bar m}$. Finally, the action of the holomorphic component $\delta$ of the scalar operator $\updelta$ on the various fields is completely determined by the requirement that it satisfies the $\N=1$ subalgebra $\delta^2=0$ and $\{\delta,\d_{\bar m}\}=\partial_{\bar m}+\gauge(A_{\bar m})$.

\subsection{$\N=2$ holomorphic superspace}

We extend       the  superspace of the $\N=1$ case into one    with complex  bosonic coordinates     $z_m$, $z_{\bar{m}}$         and  five Grassmann coordinates, one scalar $\te$, two ``holomorphic'' $\vte^{m}$ and two ``antiholomorphic'' $\vte^{\bar{m}}$ ($m,\bar{m}=1,2$). The supercharges are now given by
\be \label{susy gen} \mathbb{Q}\,\, \equiv
\frac{\partial}{\partial\te}+\vte^{m}\partial_{m}+\vte^{\bar{m}}\partial_{\bar{m}}, \hspace{10mm}
\mathbb{Q}_{m} \equiv\frac{\partial}{\partial\vte^{m}},\hspace{10mm} \mathbb{Q}_{\bar{m}} \equiv\frac{\partial}{\partial\vte^{\bar m}}
\ee
They verify
\be
\mathbb{Q}^2=0,\hspace{10mm} \{\mathbb{Q},\mathbb{Q}_{M}\}=
\partial_{M}, \hspace{10mm} \{\mathbb{Q}_{M}, \mathbb{Q}_{N}\}=0
\ee
with $M=m,{\bar m}$. 
The covariant   derivatives and their  anticommuting relations are
\bea \nabla\,\, \equiv
\frac{\partial}{\partial\te} \hspace{20mm} \nabla_{M}
\equiv\frac{\partial}{\partial\vte^{M}}-\te\partial_{M} \CR
\nabla^2=0 \hspace{10mm} \{\nabla,\nabla_{M}\}= -\partial_{M}
\hspace{10mm} \{\nabla_{M}, \nabla_{N}\}=0 \eea They anticommute with the supersymmetry generators. The gauge covariant superderivatives  are    \be \hat{\nabla} \equiv \nabla +
\mathbb{C},\quad \hat{\nabla}_{M} \equiv
\nabla_{M}+\mathbf{\Upgamma}_{M},\quad \hat{\partial}_M\equiv
\partial_M+\mathbb{A}_M \ee from which we define the   covariant
superspace  curvatures   
  \be \label{ricci}\begin{split}
&\mathbb{F}_{MN} \equiv [\hat{\partial}_M,\hat{\partial}_N] \\
&\mathbf{\Uppsi}_M \equiv [\hat{\nabla},\hat{\partial}_M] \\
&\boldsymbol{\upchi}_{MN} \equiv
[\hat{\nabla}_{M},\hat{\partial}_N]
\end{split}
\hspace{10mm}\begin{split}
&\mathbf{\Upsigma} \equiv \hat{\nabla}^2 \\
&\mathbb{L}_{M} \equiv \{\hat{\nabla}, \hat{\nabla}_{M}\}+\hat{\partial}_{M} \\
&\mathbf{\bar{\Upsigma}}_{MN} \equiv
{\ \scriptstyle \frac{1}{2} } \{\hat{\nabla}_{M},\hat{\nabla}_{N} \}
\end{split}\ee
They are subjected to Bianchi identities and  the   super-gauge transformations of the various connections  and   curvatures follow  analogously to the $\N=1$ case.

The constraints for the $\N=2$ case are
\be
\mathbb{L}_{M}=0,\quad  \mathbf{\bar{\Upsigma}}_{mn}=\mathbf{\bar{\Upsigma}}_{\bar m\bar n}=0,\quad\mathbf{\bar{\Upsigma}}_{m\bar n}=\frac{1}{2}\,g_{m\bar n}\mathbf{\bar{\Upsigma}}_{p}^{\phantom{p}p},\quad \boldsymbol{\upchi}_{\bar m n}=\frac{1}{2}\,g_{\bar m n}\boldsymbol{\upchi}_p^{\,\,p}
\ee
Their solution can be directly deduced  from  \cite{twistsp,TSSlong}, by  decomposition into holomorphic and antiholomorphic coordinates. The full physical vector supermultiplet now stands in the scalar odd connexion, which in the Wess--Zumino-like gauge is
\be
\mathbb{C}= \tilde{A}+\te(\tilde{\Phi}-\tilde{A}^2)
\ee
where
\begin{multline}
 \tilde{A}=-\vte^m A_m -\vte^{\bar m}A_{\bar m} -\frac{1}{2}\vte^m\vte^n\bar{\chi}_{mn}-\frac{1}{2}\vte^{\bar m}\vte^{\bar n}\chi_{\bar m\bar n}-\vte^m\vte_m\chi\\ +\frac{1}{2}\vte^n\vte_n(\vte^m D_m\bar{\Phi}+\vte^{\bar m}D_{\bar m}\bar{\Phi})-\vte^m\vte_m\vte^{\bar m}\vte_{\bar m} [\bar{\Phi},\eta]
\end{multline}
There is an analogous decomposition for $\tilde{\Phi}$ in \cite{twistsp}.
The general solution to the constraints is recovered by the following super-gauge-transformation
\be
e^{\boldsymbol{\upalpha}}=e^{\te(\vte^{\bar m}\partial_{\bar m}+\vte^m\partial_m)}e^{\tilde{\gamma}}e^{\te\tilde{c}}= e^{\tilde{\gamma}}
\scal{1+\te( \tilde{c}+e^{-\tilde{\gamma}}(\vte^{\bar m}\partial_{\bar m}+\vte^m\partial_m)
e^{\tilde{\gamma}})}
\ee
where $\tilde{\ga}$ and $\tilde{c}$ are
respectively commuting and anticommuting functions of $\vte^{\bar m},\vte^m$
and the coordinates $z^m,z^{\bar m}$, with the condition
$\tilde{\ga}\vert_0=0$. Transformation laws in components can then be recovered, which match those in (\ref{lois N2 d4}) and (\ref{Bianchi N2 d4}).

The action is then given by
\be \label{actionEQN2} \mathcal{S}_{SYM}^{\N=2} = \int\,{\rm d}^4\vte\, {\rm d}\te\,\trace\Bigl(\mathbb{C}\nabla\mathbb{C}+\frac{2}{3}\mathbb{C}^3
 \Bigr)\ee
To recover the previous results of the $\N=1$ super--Yang--Mills theory with matter in the adjoint representation, one first integrates (\ref{actionEQN2}) over the $\te$ variable, which gives 
\be
\mathcal{S}_{SYM}^{\N=2} = \int\,{\rm d}^4\vte\, \trace
\mathbf{\Upsigma}\mathbf{\Upsigma}
\ee
Further integration over the $\vte^m$ variables, or equivalently derivation with $\nabla_m$, yields two terms which are both invariant under the $\N=1$ scalar supersymmetry generator. In turn, they can be expressed as an integral over the full twisted $\N=1$ superspace, which yields the superspace action given in (\ref{Kithira}) with matter in the adjoint representation.

\subsection*{Acknowledgments}
We thank very much G.~Bossard for many useful discussions on the subject.
This work has been partially supported by the contract ANR (CNRS-USAR), \texttt{05-BLAN-0079-01}. A.~M. has been supported by the Swiss National Science Foundation, grant \texttt{PBSK2-119127}.


\end{document}